\documentclass[a4paper,11pt]{article}
\usepackage{latexsym}
\usepackage{amssymb}
\usepackage{amsfonts}
\usepackage{amsmath}
\usepackage{indentfirst}
\usepackage[dvips]{graphicx}
\usepackage{epsfig}
\newcommand{\cvd}{\hfill $\blacksquare$\bigskip}

\newtheorem{proposition}{Proposition}[section]
\newtheorem{example}{Example}[section]

\date{}
\author{S. Bilotta\thanks{Dipartimento di Sistemi e Informatica, Universit\`a degli Studi di Firenze, Viale
 G.B. Morgagni 65, 50134 Firenze, Italy. {\newline
 \tt \ bilotta@dsi.unifi.it,\quad elisa@dsi.unifi.it,\quad pinzani@dsi.unifi.it}} \and E. Pergola$^*$\and R. Pinzani$^*$}

\title{A new approach to cross-bifix-free sets}

\begin{document}

\maketitle

\begin{abstract}
Cross-bifix-free sets are sets of words such that no prefix of any
word is a suffix of any other word. In this paper, we introduce a
general constructive method for the sets of cross-bifix-free
binary words of fixed length. It enables us to determine a
cross-bifix-free words subset which has the property to be
non-expandable.
\end{abstract}

\section{Introduction}

In digital communication systems, synchronization is an essential
requirement to establish and maintain a connection between a
transmitter and a receiver.

%The most common method for obtaining or maintaining word or frame
%synchronization in a digital communication utilizes the insertion
%of a fixed pattern at regular intervals into the transmitted
%sequence.

Analytical approaches to the synchronization acquisition process
and methods for the construction of sequences with the best
aperiodic autocorrelation properties have been the subject of
numerous analyses in the digital transmission.

The historical engineering approach started with the introduction
of bifix. It denotes a subsequence that is both a prefix and
suffix of a longer observed sequence. Rather than to the bifix,
much attention has been devoted to a bifix-indicator, an indicator
function implying the existence of the bifix \cite{10}. Such
indicators were shown to be without equal in performing various
statistical analysis, mainly concerning the search process
\cite{3,10}

However, an analytical study of simultaneous search for a set of
sequences urged the invention of cross-bifix indicators \cite{1,2}
and, accordingly, turned attention to the sets of sequences which
avoid cross-bifixes, called cross-bifix-free sets.

In \cite{1}, the author analyzes some properties of binary words
that form a cross-bifix-free set, in particular, a general
constructing method called the kernel method is presented. This
approach leads to sets $S(n)$ of cross-bifix-free binary words, of
fixed length $n$, having cardinality
$1,1,2,3,5,8,13,21,34,55,89,144,233$ for
$n=3,4,5,6,7,8,9,10,11,12,13,14,15$ respectively.

This sequence forms a Fibonacci progression and satisfies the
recurrence relation $|S(n)|=|S(n-1)|+|S(n-2)|$ with $|S(3)|=1$ and
$|S(4)|=1$.

The problem of determining cross-bifix-free sets is also related
to several other scientific applications, for instance in
multiaccess systems, pattern matching and automata theory.

The aim of this paper is to introduce a method for the generation
of sets of cross-bifix-free binary words of fixed length based
upon the study of lattice paths on the Cartesian plane. This
approach enables us to obtain cross-bifix-free sets having greater
cardinality than the ones presented in \cite{1}.

The paper is organized as follows. In Section 2 we give some basic
definitions and notation related to the notions of bifix-free word
and cross-bifix-free set. In Section 3 we propose a method to
construct particular sets of cross-bifix-free binary words of
fixed length $n$ related to the parity of $n$. We are not able to
say if such cross-bifix-free sets have maximal cardinality on the
set of bifix-free binary words of fixed length $n$ or not.

\section{Basic definitions and notations}
Let $A$ be a finite, non-empty set called \emph{alphabet}. The
elements of $A$ are called \emph{letters}. A (finite) sequence of
letters in $A$ is called (finite) \emph{word}. Let $A^*$ denote
the monoid of all finite words over $A$ where $\varepsilon$
denotes the \emph{empty word} and $A^+ = A^* \backslash
\varepsilon$. Let $\omega$ be a word in $A^*$, then $|\omega|$
indicates the length of $\omega$ and ${|\omega|}_a$ denotes the
number of occurrences of $a$ in $\omega$, being $a \in A$. Let
$\omega=uv$ then $u$ is called \emph{prefix} of $\omega$ and $v$
is called \emph{suffix} of $\omega$. A \emph{bifix} of $\omega$ is
a subsequence of $\omega$ that is both its prefix and suffix.

A word $\omega$ of $A^+$ is said to be \emph{bifix-free} or
\emph{unbordered} \cite{7,11} if and only if no strict prefix of
$\omega$ is also a suffix of $\omega$. Therefore, $\omega$ is
bifix-free if and only if $\omega \neq uwu$, being $u$ any
necessarily non-empty word and $w$ any word. Obviously, a
necessary condition for $\omega$ to be bifix-free is that the
first and the last letters of $\omega$ must be different.
\begin{example}
In the monoid $\{0,1\}^*$, the word $111010100$ of length $n=9$ is
bifix-free, while the word $101001010$ contains two bifixes, $10$
and $1010$.
\end{example}

Let $BF_q(n)$ denote the set of all bifix-free words of length $n$
over an alphabet of fixed size $q$. The following formula for the
cardinality of $BF_q(n)$, denoted by $|BF_q(n)|$, is well-known
\cite{11}.

\begin{equation}
\label{bifixfree} \left \{
\begin{array}{lll}
|BF_q(1)|=q\\\\
|BF_q(2n+1)|=q|BF_q(2n)|\\\\
|BF_q(2n)|=q|BF_q(2n-1)|-|BF_q(n)|
\end{array}
\right.
\end{equation}

The number sequences related to this recurrence can be found in
Sloane's database of integer sequences \cite{12}: sequences
A003000 ($q=2$), A019308 ($q=3$) and A019309 ($q=4$).

Table \ref{bfree} lists the set $BF_2(n)$, $2 \leq n \leq 6$, the
last row reports the cardinality of each set.

\begin{table}[htb]
\begin{center}
\begin{tabular}{|c|c|c|c|c|c|}
\hline \hline n=2 & n=3 & n=4 & n=5 & n=6\\
\hline

10 \ 01 & 100 \ 001 & 1000 \ 0001 & 10000 \ 00001 & 100000 \ 000001 \\
   & 110 \ 011 & 1100 \ 0011 & 10100 \ 00101 & 101000 \ 000101\\
   &     & 1110 \ 0111 & 11000 \ 00011 & 101100 \ 001101\\
   &     &      & 11100 \ 00111 & 110000 \ 000011\\
   &     &      & 11010 \ 01011 & 110100 \ 001011\\
   &     &      & 11110 \ 01111 & 111000 \ 000111\\
   &     &      &       & 111100 \ 001111\\
   &     &      &       & 110010 \ 010011\\
   &     &      &       & 111010 \ 010111\\
   &     &      &       & 111110 \ 011111\\
\hline
 2 & 4 & 6 & 12 & 20\\
\hline

\end{tabular}
\end{center}
\caption{\label{bfree}The set $BF_2(n)$, $2 \leq n \leq 6$}
\end{table}

Let $q>1$ and $n>1$ be fixed. Two distinct words $\omega,\omega'
\in BF_q(n)$ are said to be \emph{cross-bifix-free} if and only if
no strict prefix of $\omega$ is also a suffix of $\omega'$ and
vice-versa.

\begin{example}\label{esemp}
The binary words $111010100$ and $110101010$ in $BF_2(9)$ are
cross-bifix-free, while the binary words $111001100$ and
$110011010$ in $BF_2(9)$ have the cross-bifix $1100$.
\end{example}

A subset of $BF_q(n)$ is said to be \emph{cross-bifix-free set} if
and only if for each $\omega, \omega'$, with $\omega \neq
\omega'$, in this set, $\omega$ and $\omega'$ are
cross-bifix-free. This set is said to be \emph{non-expandable} on
$BF_q(n)$ if and only if the set obtained by adding any other word
is not a cross-bifix-free set. A non-expandable cross-bifix-free
set on $BF_q(n)$ having maximal cardinality is called
\emph{maximal cross-bifix-free set} on $BF_q(n)$.

Each word $\omega \in BF_2(n)$ can be naturally represented as a
lattice path on the Cartesian plane, by associating a \emph{rise
step}, defined by $(1,1)$ and denoted by $x$, to each 1's in
$BF_2(n)$, and a \emph{fall step}, defined by $(1,-1)$ and denoted
by $\overline{x}$, to each 0's in $BF_2(n)$, running from $(0,0)$
to $(n,h)$, $-n<h<n$.

From now on, we will refer interchangeably to words or their
graphical representations on the Cartesian plane, that is paths.

\begin{figure}[!htb]
\begin{center}
\epsfig{file=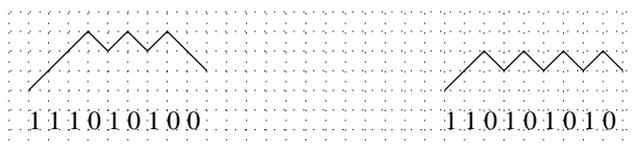,width=3.3in,clip=} \caption{\small{Two
paths in $BF_2(9)$ which are cross-bifix-free}
\label{exmpcrossbf}}\vspace{-15pt}
\end{center}
\end{figure}

The definition of bifix-free and cross-bifix-free can be easily
extended to paths. Figure \ref{exmpcrossbf} shows the two paths
corresponding to the cross-bifix-free words of Example
\ref{esemp}.

\begin{figure}[!htb]
\begin{center}
\epsfig{file=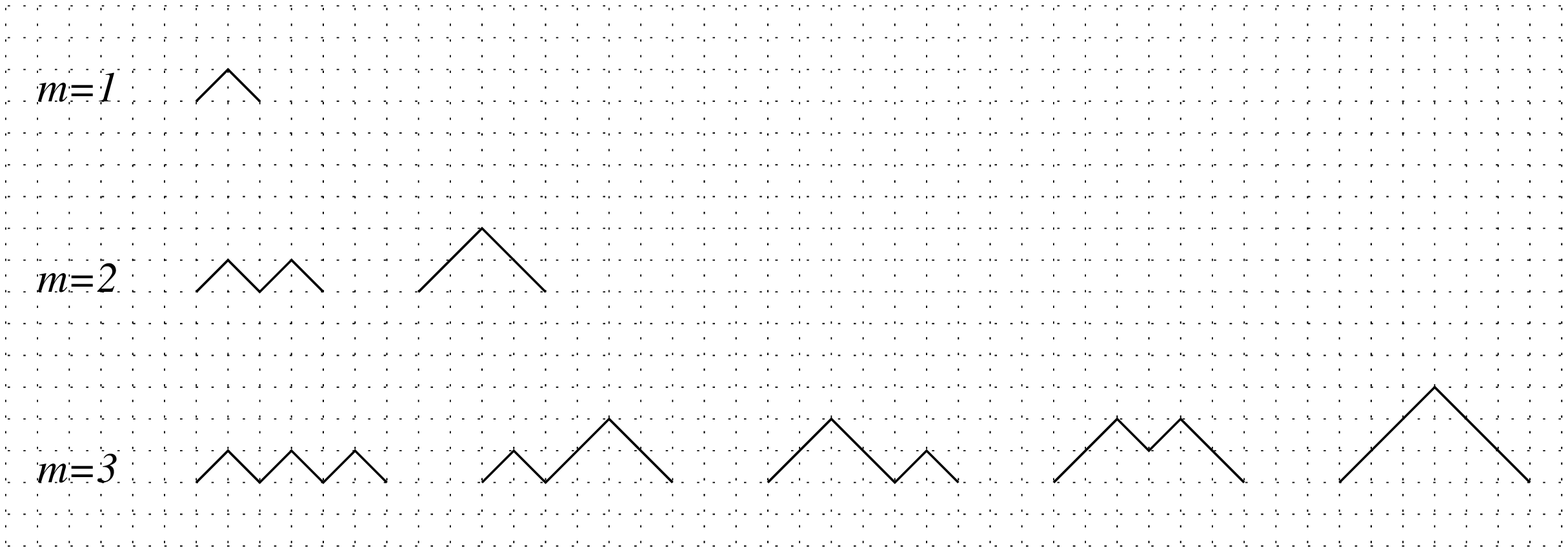,width=5in,clip=} \caption{\small{The
$2m$-length Dyck paths, $1 \leq m \leq 3$}
\label{dick}}\vspace{-15pt}
\end{center}
\end{figure}

A lattice path on the Cartesian plane using the steps $(1,1)$ and
$(1,-1)$ and running from $(0,0)$ to $(2m,0)$, with $m \geq 0$, is
said to be \emph{Grand-Dyck} or \emph{Binomial} path (see \cite{5}
for further details). A \emph{Dyck} path is a sequence of rise
step and fall steps running from $(0,0)$ to $(2m,0)$ and remaining
weakly above the $x$-axis (see Figure \ref{dick}). The number of
$2m$-length Dyck paths is the $m$th Catalan number $C_m=1/(m+1){2m
\choose m}$, see \cite{13} for further details.

In this paper, we are interested in investigating a possible
non-expandable cross-bifix-free set, that is the set $CBFS_2(n)$
of cross-bifix-free words of fixed length $n>1$ on the monoid
$\{0,1\}^*$. In order to do so, we focus on the set
$\hat{BF}_2(n)$ of bifix-free binary words of fixed length $n$
having 1 as the first letter and 0 as last letter or equivalently
the set of bifix-free lattice paths on the Cartesian plane using
the steps $(1,1)$ and $(1,-1)$, running from $(0,0)$ to $(n,h)$,
$-n<h<n$, beginning with a rise step and ending with a fall step.
Of course $\check{BF}_2(n) = BF_2(n) \backslash \hat{BF}_2(n)$ is
obtained by switching rise and fall steps.

Let $\hat{BF}_{2}^{h}(n)$ denote the set of the paths in
$\hat{BF}_2(n)$ having $h$ as the ordinate of their endpoint, $-n
< h < n$.

\section{On the non-expandability of $CBFS_2(n)$}
In order to prove that $CBFS_2(n)$ is a non-expandable
cross-bifix-free set on $BF_2(n)$ we have to distinguish the
following two cases depending on the parity of $n$.

\subsection{Non-expandable $CBFS_2(2m+1)$} 
\begin{figure}[!htb]
\begin{center}
\epsfig{file=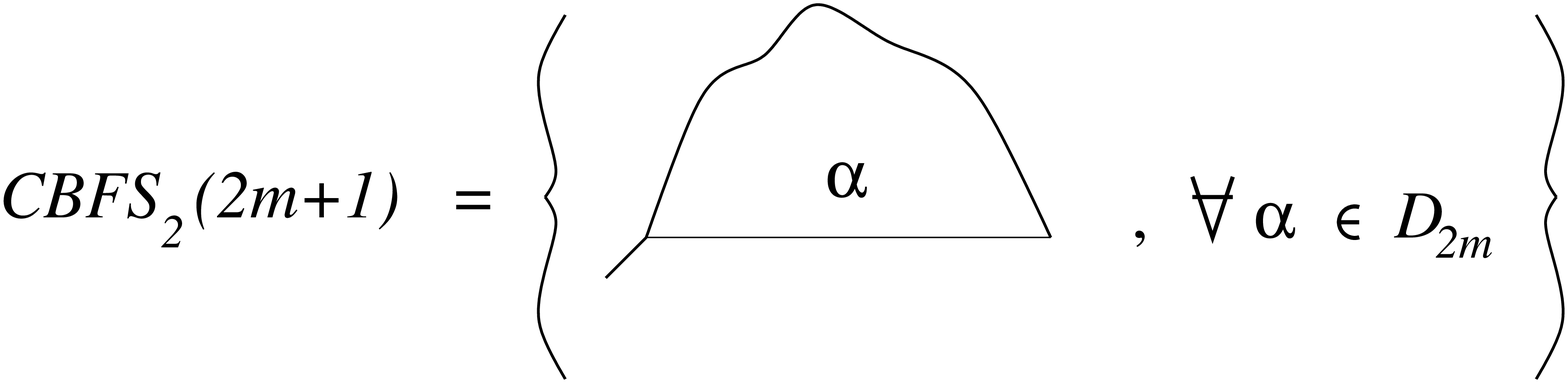,width=3.5in,clip=} \caption{\small{A
graphical representation of $CBFS_2(2m+1)$, with $m \geq 1$}
\label{rappdis}}\vspace{-15pt}
\end{center}
\end{figure}

Let $CBFS_2(2m+1) = \{x
\alpha : \alpha \in D_{2m} \}$ that is the set of paths beginning
with a rise step linked to a $2m$-length Dyck path (see Figure
\ref{rappdis}). Note that $CBFS_2(2m+1)$ is a subset of
$\hat{BF}^1_2(2m+1)$, $m \geq 1$.

\begin{figure}[!htb]
\begin{center}
\epsfig{file=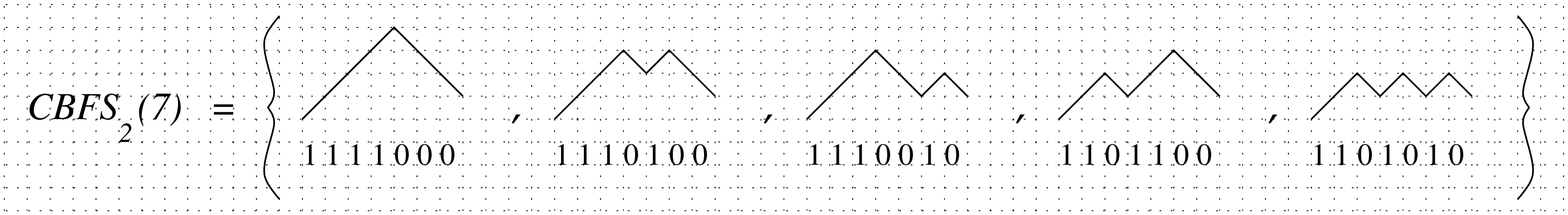,width=6in,clip=} \caption{\small{A
graphical representation of $CBFS_2(7)$}
\label{rap7}}\vspace{-15pt}
\end{center}
\end{figure}

Of course $|CBFS_2(2m+1)|=C_m$, being $C_m$ the $m$th
Catalan number, $m \geq 1$.\\
Figure \ref{rap7} shows the set $CBFS_2(7)$, with
$|CBFS_2(7)|=C_3=5$.

\begin{proposition}\label{dispariCB}
The set $CBFS_2(2m+1)$ is a cross-bifix-free set on $BF_2(2m+1)$,
$m \geq 1$.
\end{proposition}
\emph{Proof.}\quad The proof consists of two distinguished steps.
The first one proves that each $\omega \in CBFS_2(2m+1)$ is
bifix-free and the second one proves that $CBFS_2(2m+1)$ is a
cross-bifix-free set. Each $\omega \in CBFS_2(2m+1)$ can be
written as $\omega=vwu$, being $v,u$ any necessarily non-empty
word while $w$ can also be an empty word. For each prefix $v$ of
$\omega$ we have $|v|_1
> |v|_0$ and for each suffix $u$ of $\omega$ we have $|u|_1
\leq |u|_0$. Therefore $v \neq u$, $\forall v,u \in \omega$ so
$\omega$ is bifix-free.

The proof that, for each $\omega,\omega' \in CBFS_2(2m+1)$ then
$\omega$ and $\omega'$ are cross-bifix-free, follows the logical
steps described above. \cvd

\begin{proposition}\label{dispariNE}
The set $CBFS_2(2m+1)$ is a non-expandable cross-bifix-free set on
$BF_2(2m+1)$, $m \geq 1$.
\end{proposition}
\emph{Proof.}\quad It is sufficient to prove that the set
$CBFS_2(2m+1)$ is a non-expandable cross-bifix-free set on
$\hat{BF}_2(2m+1)$, as each $\omega \in CBFS_2(2m+1)$ and $\varphi
\in \check{BF}_2(2m+1)$ match on the last letter of $\omega$ and
the first one of $\varphi$ at least.

Let $m \geq 1$ be fixed, we can prove that $CBFS_2(2m+1)$ is a
non-expandable cross-bifix-free set on $\hat{BF}^h_2(2m+1)$ by
distinguishing $h>0$ from $h<0$.
\begin{itemize}
\item[$\bullet$ $h>0$ :] a path $\gamma$ in $\hat{BF}^h_2(2m+1)
\backslash CBFS_2(2m+1)$ can be written as $\gamma=\phi x \alpha_1
x \alpha_2 x \dots x \alpha_r$ (see Figure \ref{classep}, where
$n=2m+1$), being $\phi$ a Grand-Dyck path beginning with a rise
step, $x$ a rise step, $\alpha_l$ Dyck paths, $1 \leq l \leq r-1$,
and $\alpha_r$ a necessarily non-empty Dyck path. Therefore, we
can find paths in $CBFS_2(2m+1)$ having a prefix which matches
with a suffix of $\gamma$. It is sufficient to consider the path
$\omega = x \alpha_r \alpha_s$, being $\alpha_s$ a Dyck path of
appropriate length.

\begin{figure}[!htb]
\begin{center}
\epsfig{file=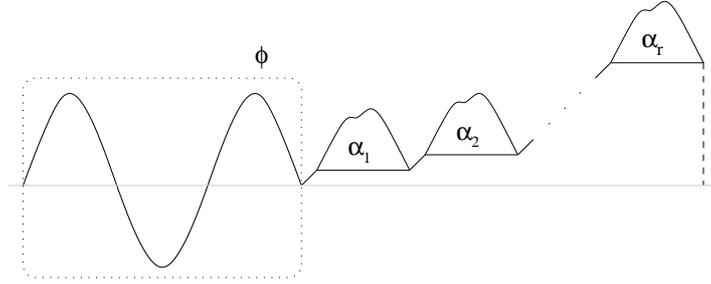,width=3.7in,clip=} \caption{\small{A
graphical representation of a path $\gamma$ in $\hat{BF}^h_2(n),
h>0$} \label{classep}}\vspace{-15pt}
\end{center}
\end{figure}

\item[$\bullet$ $h<0$ :] a path $\gamma$ in $\hat{BF}^h_2(2m+1)$
can be written as $\gamma= \alpha_r \overline{x} \alpha_{r-1}
\overline{x} \dots \overline{x} \alpha_1 \overline{x} \phi$ (see
Figure \ref{classepnegativa}, where $n=2m+1$), being $\alpha_r$ a
necessarily non-empty Dyck path, $\overline{x}$ a fall step,
$\alpha_l$ Dyck paths, $1 \leq l \leq r-1$, and $\phi$ a
Grand-Dyck path. Therefore, we can find paths in $CBFS_2(2m+1)$
having a suffix which matches with a prefix of $\gamma$. It is
sufficient to consider the path $\omega = x \alpha_s \alpha_r$,
being $\alpha_s$ a Dyck path of appropriate length.

\begin{figure}[!htb]
\begin{center}
\epsfig{file=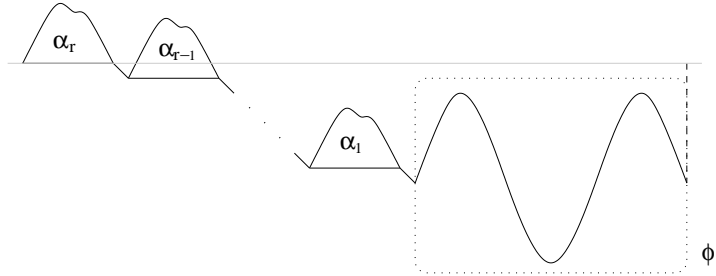,width=3.7in,clip=} \caption{\small{A
graphical representation of a path $\gamma$ in $\hat{BF}^h_2(n),
h<0$} \label{classepnegativa}}\vspace{-12pt}
\end{center}
\end{figure}
\end{itemize}
\cvd

\subsection{Non-expandable $CBFS_2(2m+2)$}
In this case we have to distinguish
two further subcases depending on the parity of $m > 0$.\\

\begin{figure}[!htb]
\begin{center}
\epsfig{file=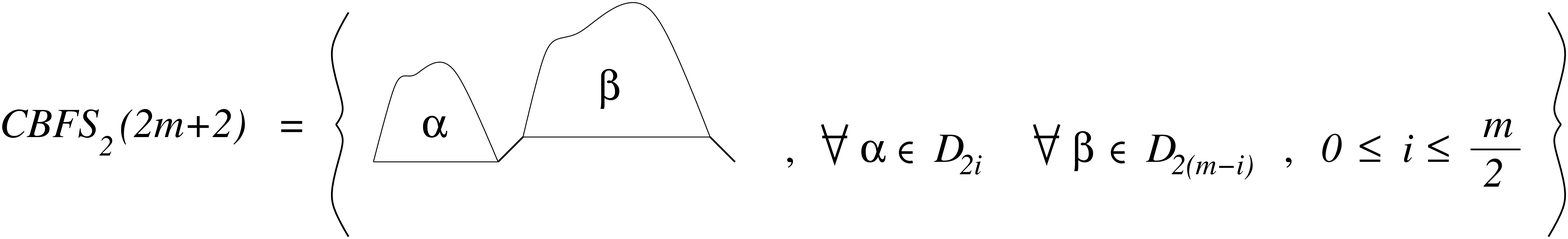,width=5.3in,clip=} \caption{\small{A
graphical representation of $CBFS_2(2m+2)$, for any even number
$m>1$} \label{paripari}}\vspace{-15pt}
\end{center}
\end{figure}

If $m$ is an even number then $CBFS_2(2m+2) = \{ \alpha x \beta
\overline{x} : \alpha \in D_{2i} \ , \ \beta \in D_{2(m-i)} \ , \
0 \leq i \leq \frac{m}{2} \}$, that is the set of paths consisting
of the following consecutive sub-paths: a $2i$-length Dyck path, a
rise step, a $2(m-i)$-length Dyck path, a fall step, where $0 \leq
i \leq \frac{m}{2}$ (see Figure \ref{paripari}). Note that
$CBFS_2(2m+2)$ is a subset of $\hat{BF}^0_2(2m+2)$, for any even
number $m>1$.

\begin{figure}[!htb]
\begin{center}
\epsfig{file=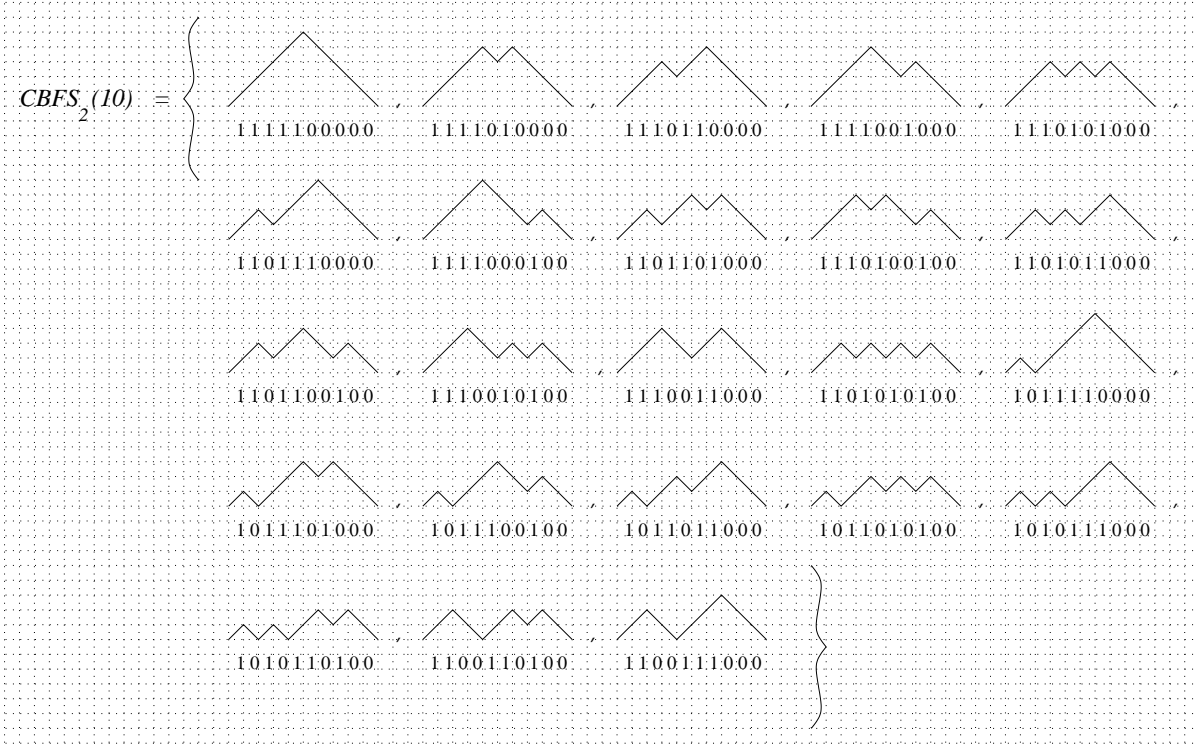,width=6.2in,clip=} \caption{\small{A
graphical representation of $CBFS_2(10)$}
\label{exmpari}}\vspace{-15pt}
\end{center}
\end{figure}

Of course $|CBFS_2(2m+2)|=\sum_{i=0}^{m/2}C_iC_{m-i}$, $C_m$ is
the $m$th Catalan Number, for any even number $m>1$. Figure
\ref{exmpari} shows the set $CBFS_2(10)$, with
$|CBFS_2(10)|=C_4+C_1C_3+C_2C_2=23$.

\begin{proposition}\label{paripariCB}
The set $CBFS_2(2m+2)$ is a cross-bifix-free set on $BF_2(2m+2)$,
for any even number $m > 1$.
\end{proposition}
\emph{Proof.}\quad The proof consists of two distinguished steps.
The first one proves that each $\omega \in CBFS_2(2m+2)$ is
bifix-free and the second one proves that $CBFS_2(2m+2)$ is a
cross-bifix-free set. Each $\omega \in CBFS_2(2m+2)$ can be
written as $\omega=vwu$, being $v,u$ any necessarily non-empty
word while $w$ can also be an empty word. Let $m > 1$ be fixed, we
have to take into consideration two different cases: in the first
one $i=0$ and in the second one $0 < i \leq \frac{m}{2}$.

If $i=0$ then $\omega \in \{x \beta \overline{x} : \beta \in
D_{2m} \}$, and for each prefix $v$ of $\omega$ we have $|v|_1
> |v|_0$ and for each suffix $u$ of $\omega$ we have $|u|_1 <
|u|_0$. Therefore $v \neq u$, $\forall v,u \in \omega$ and
$\omega$ is bifix-free.

Otherwise, $\omega \in \{ \alpha x \beta \overline{x} : \alpha \in
D_{2i} \ , \ \beta \in D_{2(m-i)} \ , \ 0 < i \leq \frac{m}{2}
\}$, then for each prefix $v$ of $\omega$ we have $|v|_1 \geq
|v|_0$ and for each suffix $u$ of $\omega$ we have $|u|_1 \leq
|u|_0$. If $|v|_1 > |v|_0$ then $v \neq u$, $\forall v,u \in
\omega$ and therefore $\omega$ is bifix-free. Let $i$ be fixed, if
$|v|_1=|v|_0$ then the path $v$ is a $2k$-length Dyck path, $1
\leq k \leq i$. In this case, both $u=\mu \overline{x}$, where
$\mu$ is any suffix of $\beta$, and $u=\mu' x \beta \overline{x}$,
where $\mu'$ is any suffix of $\alpha \backslash v$. If $u=\mu
\overline{x}$ then $|u|_1 < |u|_0$, therefore $v \neq u$, $\forall
v,u \in \omega$ and therefore $\omega$ is bifix-free. If $u=\mu' x
\beta \overline{x}$ then $v$ does not match with $x \beta
\overline{x}$, therefore $v \neq u$, $\forall v,u \in \omega$ and
therefore $\omega$ is bifix-free.

The proof that, for each $\omega,\omega' \in CBFS_2(2m+2)$ then
$\omega$ and $\omega'$ are cross-bifix-free, follows the logical
steps described above. \cvd

\begin{proposition}\label{paripariNE}
The set $CBFS_2(2m+2)$ is a non-expandable cross-bifix-free set on
$BF_2(2m+2)$, for any even number $m > 1$.
\end{proposition}
\emph{Proof.}\quad It is sufficient to prove that the set
$CBFS_2(2m+2)$ is a non-expandable cross-bifix-free set on
$\hat{BF}_2(2m+2)$, as each $\omega \in CBFS_2(2m+2)$ and $\varphi
\in \check{BF}_2(2m+2)$ match on the last letter of $\omega$ and
the first one of $\varphi$ at least.

Let $m > 1$ be fixed, we have to take into consideration three
different cases: in the first one we prove that $CBFS_2(2m+2)$ is
a non-expandable cross-bifix-free set on $\hat{BF}^h_2(2m+2)$,
$h>0$, in the second one we prove that $CBFS_2(2m+2)$ is a
non-expandable cross-bifix-free set on $\hat{BF}^h_2(2m+2)$,
$h<0$, and in the last one we prove that $CBFS_2(2m+2)$ is a
non-expandable cross-bifix-free set on $\hat{BF}^0_2(2m+2)$.

\begin{itemize}
\item[$\bullet$ $h>0$ :] a path $\gamma$ in $\hat{BF}^h_2(2m+2)$
can be written as $\gamma=\phi x \alpha_1 x \alpha_2 x \dots x
\alpha_r$ (see Figure \ref{classep}, where $n=2m+2$), being $\phi$
a Grand-Dyck path beginning with a rise step, $x$ a rise step,
$\alpha_l$ Dyck paths, $1 \leq l \leq r-1$, and $\alpha_r$ a
necessarily non-empty Dyck path. Therefore, we can find paths in
$CBFS_2(2m+2)$ having a prefix which matches with a suffix of
$\gamma$. It is sufficient to consider the path $\omega = x
\alpha_r \alpha_s \overline{x}$, being $\alpha_s$ a Dyck path of
appropriate length.

\item[$\bullet$ $h<0$ :] a path $\gamma$ in $\hat{BF}^h_2(2m+2)$
can be written as $\gamma= \alpha_r \overline{x} \alpha_{r-1}
\overline{x} \dots \overline{x} \alpha_1 \overline{x} \phi$ (see
Figure \ref{classepnegativa}, where $n=2m+2$), being $\alpha_r$ a
necessarily non-empty Dyck path, $\overline{x}$ a fall step,
$\alpha_l$ Dyck paths, $1 \leq l \leq r-1$, and $\phi$ a
Grand-Dyck path. Therefore, we can find paths in $CBFS_2(2m+2)$
having a suffix which matches with a prefix of $\gamma$. It is
sufficient to consider the path $\omega = x \alpha_s \alpha_r
\overline{x}$, being $\alpha_s$ a Dyck path of appropriate length.

\item[$\bullet$ $h=0$ :] a path $\gamma$ in $\hat{BF}^0_2(2m+2)
\backslash CBFS_2(2m+2)$ either never falls below the $x$-axis or
crosses the $x$-axis. In the first case, it can be written as
$\gamma = \alpha_1 x \beta_1 \overline{x}$, where $\alpha_1$ is a
necessarily non-empty $2k$-length Dyck path and $\beta_1$ is a
$2(m-k)$-length Dyck path, with $\frac{m}{2} + 1 \leq k \leq m$,
see Figure \ref{parparNE} a). Therefore, we can find paths in
$CBFS_2(2m+2)$ having a prefix which matches with a suffix of
$\gamma$. It is sufficient to consider the path $\omega = x
\beta_1 \overline{x} x \beta \overline{x}$, since $x \beta_1
\overline{x} \in D_{2i}$ being $i=m-k+1$.

If a path $\gamma$ in $\hat{BF}^0_2(2m+2) \backslash CBFS_2(2m+2)$
crosses the $x$-axis then it can be written as $\gamma = \alpha_1
\phi$ where $\alpha_1$ is a necessarily non-empty $2k$-length Dyck
path, $1 \leq k \leq m$, and $\phi$ is a necessarily non-empty
Grand-Dyck beginning with a fall step, see Figure \ref{parparNE}
b). Therefore, we can find paths in $CBFS_2(2m+2)$ having a suffix
which matches with a prefix of $\gamma$. It is sufficient to
consider the path $\omega = x \alpha_s \alpha_1 \overline{x}$,
being $\alpha_s$ a Dyck path of appropriate length.

\begin{figure}[!htb]
\begin{center}
\epsfig{file=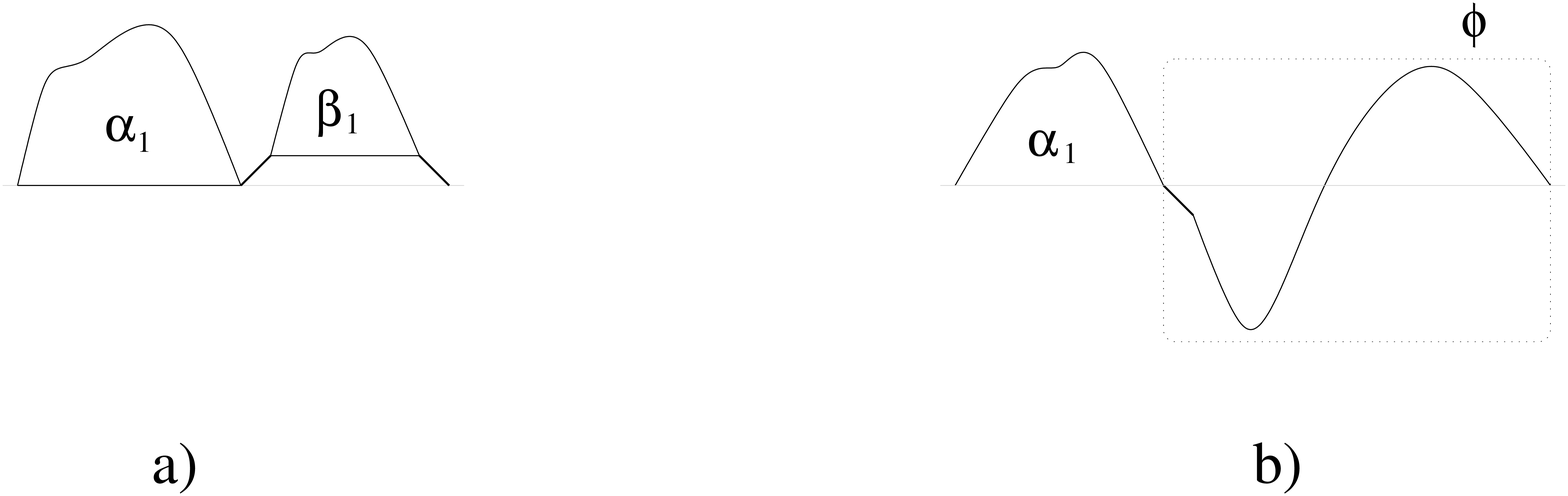,width=4.5in,clip=} \caption{\small{The two
possible configurations for a path $\gamma$ in $\hat{BF}^0_2(2m+2)
\backslash CBFS_2(2m+2)$, for any even number $m>1$}
\label{parparNE}}\vspace{-12pt}
\end{center}
\end{figure}
\end{itemize}
\cvd

\begin{figure}[!htb]
\begin{center}
\epsfig{file=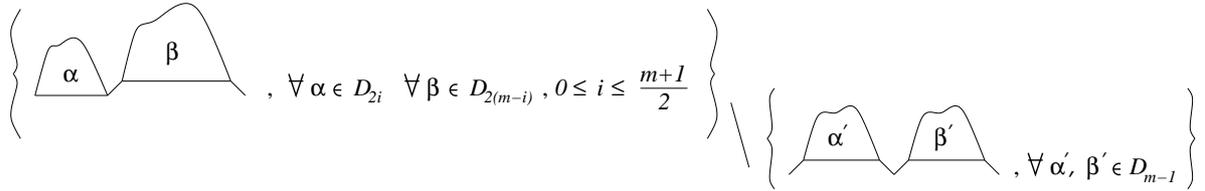,width=6.2in,clip=}
\caption{\small{A graphical representation of $CBFS_2(2m+2)$, for
any odd number $m \geq 1$} \label{paridispari}}\vspace{-15pt}
\end{center}
\end{figure}

If $m$ is an odd number then $CBFS_2(2m+2) = \{ \alpha x \beta
\overline{x} : \alpha \in D_{2i} \ , \ \beta \in D_{2(m-i)} \ , \
0 \leq i \leq \frac{m+1}{2} \} \backslash \{ x \alpha'
\overline{x} x \beta' \overline{x} : \alpha', \beta' \in D_{m-1}
\}$, that is the set of paths consisting of the following
consecutive sub-paths: a $2i$-length Dyck path, a rise step, a
$2(m-i)$-length Dych path, a fall step, where $0 \leq i \leq
\frac{m+1}{2}$, and excluding those consisting of the following
consecutive sub-paths: a rise step, a $(m-1)$-length Dyck path, a
fall step followed by a rise step, a $(m-1)$-length Dyck path, a
fall step (see Figure \ref{paridispari}). In other words, the
paths which result from the concatenation of two elevated Dyck
paths of the same length must be excluded.

In particular, if $\alpha'=\beta'$ then the excluded paths are not
bifix-free, otherwise if $\alpha' \neq \beta'$ then the excluded
paths match with the paths $\{ \alpha x \beta \overline{x} :
\alpha \in D_{m+1} \ , \ \beta \in D_{2(m-1)} \}$ in
$CBFS_2(2m+2)$. Note that $CBFS_2(2m+2)$ is a subset of
$\hat{BF}^0_2(2m+2)$, for any odd number $m \geq 1$.

\begin{figure}[!htb]
\begin{center}
\epsfig{file=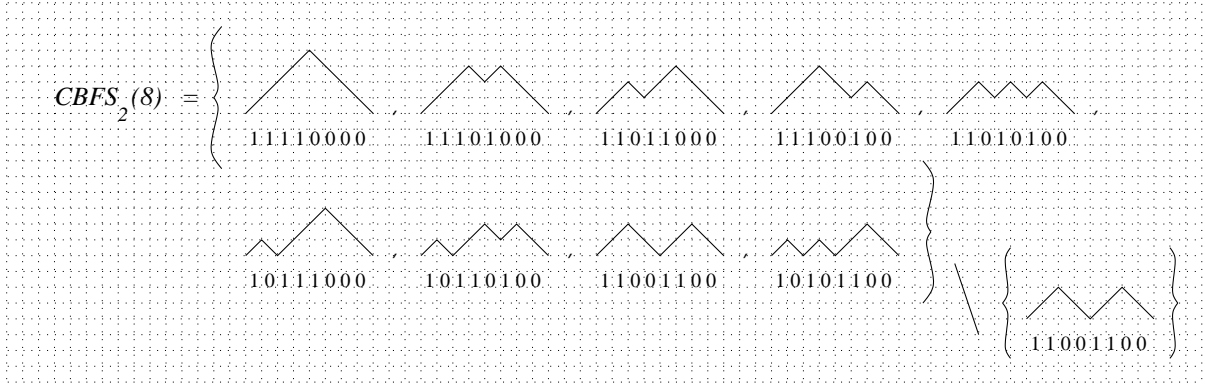,width=6.2in,clip=} \caption{\small{A
graphical representation of the set $CBFS_2(8)$}
\label{exmparidis}}\vspace{-15pt}
\end{center}
\end{figure}

Of course $|CBFS_2(2m+2)|=(\sum_{i=0}^{\frac{m+1}{2}}C_iC_{m-i}) -
(C_{\frac{m-1}{2}})^2$, $C_m$ is the $m$th Catalan Number, for any
odd number $m \geq 1$. Figure \ref{exmparidis} shows the set
$CBFS_2(8)$, with $|CBFS_2(8)|=(C_3+C_1C_2+C_2C_1) - (C_1)^2 = 8$.

\begin{proposition}\label{paridispariCB}
The set $CBFS_2(2m+2)$ is a cross-bifix-free set on $BF_2(2m+2)$,
for any odd number $m \geq 1$.
\end{proposition}

\begin{proposition}\label{padispariNE}
The set $CBFS_2(2m+2)$ is a non-expandable cross-bifix-free set on
$BF_2(2m+2)$, for any odd number $m \geq 1$.
\end{proposition}

The proof of Proposition \ref{paridispariCB} follows the logical
steps as far Proposition \ref{paripariCB} and the proof of
Proposition \ref{padispariNE} follows the logical steps as far
Proposition \ref{paripariNE}.

Therefore, the presented constructing method gives sets
$CBFS_2(n)$ of cross-bifix-free binary words, of fixed length $n$,
having cardinality $1,1,2,3,5,8,14,23,42,72,132,227,429$ for
$n=~3,4,5,6,7,8,9,10,11,12,13,14,15$ respectively.

\section{Conclusions and further developments}
In this paper, we introduce a general constructing method for the
sets of cross-bifix-free binary words of fixed length $n$ based
upon the study of lattice paths on the Cartesian plane. This
approach enables us to obtain the cross-bifix-free set $CBFS_2(n)$
having greater cardinality than the ones presented in \cite{1}
based upon the kernel method.

Moreover, we prove that $CBFS_2(n)$ is a non-expandable
cross-bifix-free set on $BF_2(n)$, i.e. $CBFS_2(n) \cup \gamma$ is
not a cross-bifix-free set on $BF_2(n)$, for any $\gamma \in
BF_2(n) \backslash CBFS_2(n)$. The non-expandable property is
obviously a necessary condition to obtain a maximal
cross-bifix-free set on $BF_2(n)$, anyway we are not able to find
and prove a sufficient condition.

Further studies to prove that could investigate both the
nontrivial subsets of $BF_2(n)$ in which $CBFS_2(n)$ is a maximal
cross-bifix-free set, and the study of other non-expandable
cross-bifix-free sets on $BF_2(n)$.

Another approach to reach the goal could be to find a different
characterization of bifix-free words which could be obtained
through bijective methods between particular bifix-free subsets
and other well-known discrete structures.

Successive studies should take into consideration the general
study of cross-bifix-free sets on $BF_q(n)$, where $q$ is grater
than 2.

\end{document}